\documentclass[a4paper,12pt, epsfig]{article}
\pdfoutput=1
\usepackage{epsfig}
\usepackage{epstopdf}
\usepackage{graphicx}
\usepackage{ifthen}
\usepackage{feynmp}
\usepackage{amsmath}
\usepackage[psamsfonts]{amssymb}
\usepackage{euscript}

\usepackage{latexsym}
\usepackage[arrow,matrix,curve]{xy}

\newskip\humongous \humongous=0pt plus 1000pt minus 1000pt

\newif\ifdtup

\def\,{\hspace{-.1cm}}
\def\hsp{,\hspace{.7cm}}

\def\fc#1#2 {\frac{n}{q}#1\frac{n}{q}#2}

\def\expb#1{{\rm{Exp}}\left[#1\right]}
\renewcommand{\cos}{\textrm{cos}}

\renewcommand{\theequation}{\arabic{section}.\arabic{equation}}
\renewcommand{\(}{\begin{equation}}
\renewcommand{\)}{end{equation} \vspace{-.05in}\linebreak}

\newcounter{saveeqn}
\newcounter{savealpheqn}

\newcommand{\alpheqn}{\setcounter{saveeqn}{\value{equation}}%
  \stepcounter{saveeqn}\setcounter{equation}{0}%
  \renewcommand{\theequation}{\mbox{\arabic{section}.\arabic{saveeqn}
\alph{equation}}}
  \renewcommand{\)}{\end{equation}}}
\def\part#1{\frac{\partial}{\partial{#1}}}%
\def\group#1{\refstepcounter{equation}\setcounter{saveeqn}
 {\value{equation}}%
  \label{#1}\setcounter{equation}{0}%
\renewcommand{\theequation}{\mbox{\arabic{section}.\arabic{saveeqn}
\alph{equation}}}
  \renewcommand{\)}{\end{equation}}}
\newcommand{\reseteqn}{\setcounter{equation}{\value{saveeqn}}%
  \renewcommand{\theequation}{\arabic{section}.\arabic{equation}}%
  \renewcommand{\)}{\end{equation}}}

\newcommand{\aalpheqn}{\setcounter{saveeqn}{\value{equation}}%
  \stepcounter{saveeqn}\setcounter{equation}{0}%
  \renewcommand{\theequation}{\mbox{
        \Alph{subsection}.\arabic{saveeqn}\alph{equation}}}
   \renewcommand{\)}{\end{equation}}}
\newcommand{\areseteqn}{\setcounter{equation}{\value{saveeqn}}%
  \renewcommand{\theequation}{\Alph{subsection}.\arabic{equation}}%
  \renewcommand{\)}{\end{equation}}}

\renewcommand{\thefootnote}{\alph{footnote}}
\renewcommand{\(}{\begin{equation}}
\renewcommand{\)}{\end{equation}}
\newcommand{\ba}{\begin{eqnarray}}
\newcommand{\ea}{\end{eqnarray}}
\newcommand{\cbp}{\mathop{\vtop{\ialign{##\crcr
   $\hfil\displaystyle{}\hfil$\crcr\noalign{\kern-13pt\nointerlineskip}
   \BIG{)}\hskip0pt\crcr\noalign{\kern3pt}}}}}
\newcommand{\pa}{\mathop{\vtop{\ialign{##\crcr
    
$\hfil\displaystyle{\oplus}\hfil$\crcr\noalign{\kern+1pt\nointerlineskip 
}
   \hspace{.08in}$^{\alpha=0}$\hskip6pt\crcr\noalign{\kern3pt}}}}}
\renewcommand{\hsp}{,\hspace{.3in}}
\newcommand{\p}{^\prime}

\def\Re#1{{\rm{Re}}\left(#1\right)}

\catcode`\@=11
\def\vereq#1#2{\lower3pt\vbox{\baselineskip1.5pt \lineskip1.5pt
\ialign{$\m@th#1\hfill##\hfil$\crcr#2\crcr\sim\crcr}}}
\catcode`\@=12

\renewcommand{\(}{\begin{equation}}
\renewcommand{\)}{\end{equation}}

\newcommand{\E}{{\mathcal{E}}}
\newcommand{\beas}{\begin{eqnarray*}}
\newcommand{\eeas}{\end{eqnarray*}}

\newcommand{\bquo}{\begin{quote}}
\newcommand{\enqu}{\end{quote}}

\newcommand{\beq}{\begin{equation}}
\newcommand{\eeq}{\end{equation}}
\newcommand{\bea}{\begin{eqnarray}}
\newcommand{\eea}{\end{eqnarray}}

\newskip\humongous \humongous=0pt plus 1000pt minus 1000pt

\newif\ifdtup

\catcode`\@=11

\@addtoreset{equation}{section}

\def\@normalsize{\@setsize\normalsize{15pt}\xiipt\@xiipt
\abovedisplayskip 14pt plus3pt minus3pt%
\belowdisplayskip \abovedisplayskip
\abovedisplayshortskip \z@ plus3pt%
\belowdisplayshortskip 7pt plus3.5pt minus0pt}

\def\small{\@setsize\small{13.6pt}\xipt\@xipt
\abovedisplayskip 13pt plus3pt minus3pt%
\belowdisplayskip \abovedisplayskip
\abovedisplayshortskip \z@ plus3pt%
\belowdisplayshortskip 7pt plus3.5pt minus0pt
\def\@listi{\parsep 4.5pt plus 2pt minus 1pt
      \itemsep \parsep
      \topsep 9pt plus 3pt minus 3pt}}

\relax

\catcode`@=12

\catcode`\@=11

\def\section{\@startsection{section}{1}{\z@}{3.5ex plus 1ex minus  .2ex}{2.3ex plus .2ex}{\large\bf}}

\def\thesection{\arabic{section}}
\def\thesubsection{\arabic{section}.\arabic{subsection}}

\def\appendix{\setcounter{section}{0}
 \def\thesection{Appendix \Alph{section}}
 \def\thesubsection{\Alph{section}.\arabic{subsection}}
 \def\theequation{\Alph{section}.\arabic{equation}}}
\renewcommand{\theequation}{\arabic{section}.\arabic{equation}}

\begin{document}
% ======================================================================== 
\def\thefootnote{\fnsymbol{footnote}}
\def\thetitle{Approximate Neutrino Oscillations in the Vacuum}
\def\autthree{Hosam Mohammed}
\def\auttwo{Jarah Evslin}
\def\autone{Emilio Ciuffoli}
\def\affa{Institute of Modern Physics, NanChangLu 509, Lanzhou 730000, China}
\def\affb{University of the Chinese Academy of Sciences, YuQuanLu 19A, Beijing 100049, China}

\begin{center}
{\large {\bf \thetitle}}

\bigskip

\bigskip

{\large \noindent  \autone{${}^{1}$}\footnote{emilio@impcas.ac.cn}, \auttwo{${}^{1,2}$}\footnote{jarah@impcas.ac.cn} and \autthree{${}^{1,2}$}\footnote{hosam@impcas.ac.cn}} %and \auttwo{${}^3$} \footnote{baiyang.zhang@wigner.mta.hu } }

\vskip.7cm

1) \affa\\
2) \affb\\
%3) \affc\\

\end{center}

\begin{abstract}
\noindent
It is well known that neutrino oscillations may damp due to decoherence caused by the separation of mass eigenstate wave packets or by a baseline uncertainty of order the oscillation wave length.  In this note we show that if the particles created together with the neutrino are not measured and do not interact with the environment, then the first source of decoherence is not present.  This demonstration uses the saddle point approximation and also assumes that the experiment lasts longer than a certain threshold.     We independently derive this result using the external wave packet model and also using a model in which the fields responsible for neutrino production and detection are treated dynamically.  Intuitively this result is a consequence of the fact that the neutrino emission time does not affect the final state and so amplitudes corresponding to distinct emission times must be added coherently.  This fact also implies that oscillations resulting from mass eigenstates which are detected simultaneously arise from neutrinos which were not created simultaneously but are nonetheless coherent, realizing the neutrino oscillation paradigm of Kobach, Manohar and McGreevy.

\end{abstract}

% \vfill
% 
% \end{titlepage}
\setcounter{footnote}{0}
\renewcommand{\thefootnote}{\arabic{footnote}}

% \pacs{??}

%\ifthenelse{\equal{\pr}{1}}{
%\maketitle
%}{}

%\pr
%\ifthenelse{\equal{\pr}{1}}{yippy}

\section{Introduction}

Surprisingly, neutrino oscillations in the vacuum have received less attention than those in a medium.  This is because interactions of the particles responsible for neutrino production with the environment are implicit in standard treatments in a sense that we will now review.  Historically, the first neutrino oscillation calculations treated neutrinos as monochromatic plane waves.  Clearly this approach is inconsistent as plane waves are homogeneous and so oscillation minima and maxima will be superimposed.   However it was eventually understood that the plane waves are just a pneumonic for a true description in terms of wave packets, which are no longer monochromatic and so may be spatially localized.  The spatial localization allows for oscillations and decoherence as desired.  But how does the spatial localization arise?

The spatial localization of a neutrino wave packet is equivalent to a space-time localization of the neutrino production \cite{as10}.  This can be done if the trajectory of the source particle, say a pion, is known as well the trajectory of the other particles produced along with the neutrino, such as the muon.  In practice, these final state source particles are never measured, so how does the localization arise?  The answer is given in Refs.~\cite{giunti97,giunti02}.  The source final particles need not be measured by the experimenter, it is sufficient that they interact with the environment.  If they interact with the environment, then the final state of the environment will depend upon the neutrino emission time, and so neutrinos emitted at different times can not interfere\footnote{Indeed, following Ref.~\cite{zurek}, distinct emission times will belong to distinct superselection sectors.} and neutrino oscillations are washed out, as described in Refs.~\cite{nuss76,revival}.  Localized wave packets are then simply a proxy for the fact neutrinos produced at different times cannot interfere with each other as a result of environmental interactions.  More precisely, the calculation of the neutrino probability reduces to a sum over distinct final environmental states and in each final state the neutrino wave packet is indeed localized.  Conversely, in the absence of an environment, neutrinos are not localized into wave packets, as was found in Ref.~\cite{grimus98}.

This makes for a satisfying picture, except for two questions.  First, what happens in the vacuum, when there is no environment with which to interact?  Second, in practice one knows the environmental interactions but the calculation requires the size of the wave packet.  How does one convert the environmental interactions into a wave packet size?  There have been many estimates regarding the second question \cite{nuss76,wilczek,rich} but derivations only exist in the simplest of cases \cite{accdec}.  In fact, in the case of reactor neutrinos it is still a matter of debate whether the relevant interactions are atomic or nuclear, leading to estimates which differ by many orders of magnitude.  Our long term goal is to derive a systematic approach to answer the second question.  

We feel that to approach the second question, one needs to first answer the first question.  That is the goal of the present work.  To be certain of our response, we use multiple independent derivations.  Actually, in the context of nonrelativistic quantum mechanics, the first question question has already been answered in Ref.~\cite{rich}.  However, as has been reviewed in Ref.~\cite{beuthe}, there are a number of reasons to not trust a quantum mechanical treatment of such an ultrarelativistic system.  That said, our results will in fact agree with those of Ref.~\cite{rich}.

We will first treat this problem using an external wave packet model \cite{giunti93} in Sec.~\ref{extsez}.  We will need the slightly more sophisticated model of Ref.~\cite{giunti97} because our source and detector are treated unevenly, as the first is not measured but the second is.  In this model, one considers an experiment which lasts for an infinite time.  In an infinite time, the source and detector would spread to infinity, unless additional interactions are introduced to stabilize them.  Furthermore the source, being unstable, would disappear.  Therefore in such an approach the source and detector are not treated as dynamical fields, but rather as rigid external sources which do not spread, back react, dissipate or entangle.  Given the relevance of entanglement between the source and the neutrino to oscillation physics which has been highlighted in Ref.~\cite{cgl} and also the relevance of the finite time nature of such experiments which may in principle invalidate an S-matrix treatment as was claimed in Ref~\cite{tempo}, one may also not trust the results of the external wave packet model.  Motivated by these concerns, in Sec.~\ref{dynsez} we provide a manifestly finite time treatment of the problem in quantum field theory, treating all fields responsible for production and detection as fully dynamical quantum fields, using the model introduced in Ref.~\cite{noimisura}.  In Sec.~\ref{maxsez} we apply our result to the old question of whether there exists a maximum coherence length beyond which neutrino oscillations become unobservable.

While this manuscript was in preparation, Ref.~\cite{grimus} appeared which also considers neutrino oscillations in the vacuum, following Refs.~\cite{gs,grimus98}.  These papers, like us, are concerned with neutrino oscillations in the vacuum and also do not find any intrinsic decoherence effect due to a separation of mass eigenstates.  Curiously, Refs.~\cite{gs,grimus} also do not find decoherence due to uncertainty in the distance travelled by a neutrino, perhaps as a result of their localization of the source and detector, but this source of decoherence does appear in Ref.~\cite{grimus98}.

\section{External Wave Packet Model} \label{extsez}

\subsection{The Setup}

In this section we will treat neutrino oscillations in the vacuum using the approach of Ref.~\cite{giunti97}.  That paper uses electroweak interactions to create and destroy the neutrino inside of a rigid, but moving, external source. The computation is in 3+1 dimensions and extends over an infinite time.  Neutrinos are created in the process
\beq
\phi_{SI}\longrightarrow \phi_{SF}+l^++\nu \label{crea}
\eeq
where $\phi_{SI}$ and $\phi_{SF}$ are the initial and final source particles, for example, $\phi_{SI}$ may be a nucleus which $\beta^+$ decays.  Here $l^+$ is a charged lepton.  The neutrino is detected in the process
\beq
\nu+\phi_{DI}\longrightarrow \phi_{DF}+l^- \label{dest}
\eeq
where $\phi_{DI}$ and $\phi_{DF}$ are the initial and final source particles, for example $\phi_{DI}$ may be a free proton and $\phi_{DF}$ a neutron.  It is assumed that the particles $\phi_{SF}$, $\phi_{DF}$ and the charged leptons are observed, either by the experimenter or by the environment.  This observation is incorporated into the calculation by imposing that these particles are described by wave functions whose width is the precision with which their positions are measured and whose velocities are determined by the measurement.  These wave functions extend through all time, with fixed width, velocity and normalization.  

The neutrino is created in the overlap of the external particles in Eq.~(\ref{crea}) and is destroyed in the intersection of the external particles in Eq.~(\ref{dest}).  The amplitude is calculated by folding the neutrino propagator into these intersections.  Each wave packet has a center at each moment in time.  The centers of the production (detection) particles intersect at the average location of production (detection) in space-time.  The time difference between these two space time points is called $T$ and represents the average neutrino propagation time, while the spatial dispacement is given by the 3-vector $L$.

Following Ref.~\cite{giunti97}, the uncertainty $\sigma_S$ in the production point is given by
\beq
\frac{1}{\sigma_{S}^2}=\frac{1}{\sigma_{SI}^2}+\frac{1}{\sigma_{SF}^2}+\frac{1}{\sigma_{+}^2}
\eeq
where $\sigma_{SI}$, $\sigma_{SF}$ and $\sigma_+$ are the widths of the wave packets of $\phi_{SI}$, $\phi_{SF}$ and $l^+$ respectively.  Their velocities are denoted similarly by $v_{SI}$, $v_{SF}$ and $v_+$.  The author also defines uncertainty-weighted moments of the velocity
\beq
\langle v^k\rangle=\sigma_d^2\left(\frac{v_{SI}^k}{\sigma_{SI}^2}+\frac{v_{SF}^k}{\sigma_{SF}^2}+\frac{v_{+}^k}{\sigma_{+}^2}\right).
\eeq
Here $v$ is a 3-vector and its square is computed using the dot product.  The variance in the wave packet velocity is
\beq
\lambda_S=\langle v^2\rangle-\langle v\rangle^2.
\eeq

In all neutrino experiments of which we are aware, the particles $\phi_{SF}$ and $l^+$ are not observed.  However they do interact with the environment.  If we perform our experiment in the vacuum, so that there is no environment, then their positions will be unconstrained and so
\beq
\sigma_{SF}=\sigma_+=\infty\label{noi}
\eeq
which implies
\beq
\sigma_d=\sigma_{SI}\hsp \langle v^k\rangle=v_{SI}^k\hsp \lambda_S=0.
\eeq
In Ref.~\cite{giunti97}, the authors write that this case ``corresponds to a different physical process from the one under consideration, which can be discussed modifying the calculation presented here in the appropriate way.''  Our goal in this section is to do just this.

\subsection{Working in a Vacuum}

For brevity, we will not repeat the computation in Ref.~\cite{giunti97}, but will describe where $\lambda_S=0$ enters.  In Eq.~(13) of Ref.~\cite{giunti97} the authors present their final formula for the amplitude and find that the contribution from each neutrino mass eigenstate $i$ is equal to
\beq
A_i = C {\rm{exp}}\left[-\frac{(|L|-v_i T)^2}{2v_i^2\Omega_i}\right]\hsp
\Omega_i=\frac{2\sigma^2_S\left(v_i-L\cdot \langle v\rangle /|L|\right)^2}{v_i^2\lambda_S}+F \label{ga}
\eeq
where $v_i$ is the expected velocity of the neutrino $\nu_i$ and $F$ is finite when $\lambda_S= 0$.  The term $C$ is independent of $T$.  Note that the expression $v_i-L\cdot\langle v\rangle/|L|$ does not vanish without infinite fine tuning, even at $\lambda_S=0$, as $v_i$ depends on the neutrino masses but the other terms do not.  Therefore
\beq
\stackrel{lim}{{}_{\lambda_S\rightarrow 0}}\Omega_i=\infty
\eeq
and so in our case $A_i$ is independent of $T$.  This is in fact necessary for the consistency of our calculation, because the expected production time is not well defined when the produced particles are not measured, therefore $T$ is the center of a homogeneous distribution, which is arbitrary.

Next, following the logic of Ref.~\cite{giunti93}, one sums the amplitude over the neutrino mass eigenstates and takes the absolute value squared to obtain a probability.  The probability is then integrated over the unmeasured quantity $T$.  The only dependence on $T$ arose from the term in Eq.~(\ref{ga}).  In Ref.~\cite{giunti97} the integration of this term led to the following term in the probability, given in their Eq.~(17)
\beq
P\sim \rm{exp}\left[-\frac{L^2}{2} \frac{(v_i-v_j)^2}{v_i^2 v_j^2(\Omega_i+\Omega_j)}\right] \label{gp}
\eeq
where $i$ and $j$ are neutrino mass eigenstate indices which must be summed over.  In our case, the amplitude is independent of $T$ and so the integration over $T$ must simply yield an infinite constant, which can be normalized as usual by considering a production rate.   This result is in fact consistent with Eq.~(\ref{gp}) because in our case $\Omega_i=\Omega_j=\infty$ and so the term shown is unity.

The term in Eq.~(\ref{gp}) is the only term in the exponential which is proportional to $L^2$.  The coherence length $L_{coh}$ of neutrino oscillations is defined by the proportionality of the oscillation probability
\beq
P\propto {\rm{Exp}}\left[-\frac{L^2}{L_{coh}^2}\right].
\eeq
In our case there is no $L^2$ term in the exponential, and so the coherence length is infinite.  This is our main result.  Formally it may be obtained from Eq.~(23) of \cite{giunti97} by noting that their $\omega$ is infinite.

This is not to say that there is no decoherence.  Decoherence due to the uncertainty in the baseline does not arise from the $T$ dependent terms and in fact it is independent of both $L$ and $T$, and so it persists even in this case.  Indeed, in our case (\ref{noi}) the uncertainty in the location of the production and detection point, and so the distance travelled by the neutrino, is even larger and so one expects more decoherence in the vacuum.

\section{Fully Dynamical Model} \label{dynsez}

\subsection{Review of Analytic Results}

We will now present a second, independent derivation of this result using the model of Ref.~\cite{noimisura}.  In this model, neutrinos in the flavor eigenstate $i$ are produced by a two body decay
\beq
\phi_{SH}\longrightarrow \phi_{SL}+\psi_i 
\eeq
where $\phi_{SH}$ and $\phi_{SL}$ are the initial and final source particles, for example, $\phi_{SH}$ may be a pion and $\phi_{SL}$ an antimuon.   We will call the particle $\psi_i$ a neutrino of mass eigenstate $i$.  It is detected in the process
\beq
\psi_i+\phi_{DL}\longrightarrow \phi_{DH}.
\eeq
The indices $H$ and $L$ denote in each case the heavier and lighter particle respectively.  The fields of all particles will be evolved consistently in the Schrodinger picture of quantum field theory, using the Hamiltonian $H$, for a fixed time $t$.   We remind the reader that in the Schrodinger picture, $H$ is time-independent.

We will keep track of the full quantum states, together with all entanglements.  To keep such a computation tractable, instead of electroweak interactions we use a simplified scalar model in 1+1 dimensions.  In the ultrarelativistic limit, neutrinos may be approximated by scalars as described in Ref.~\cite{beuthe}.   Despite this brutal approximation, we will continue to refer to the field $\psi$ as a neutrino.  The Hamiltonian $H$ is the sum of the standard free massive scalar Hamiltonian $H_0$ and an interaction term
\beq
H_I=\int dx :\mathcal{H}_I(x):\hsp
\mathcal{H}_I(x)=\sum_{\alpha=\{S,D\}}\phi_{\alpha H}(x)\phi_{\alpha L}(x)\left(\psi_1(x)+\psi_2(x)\right). 
\eeq
The sum $\psi_1+\psi_2$ represents a flavor eigenstate and colons denote the standard normal ordering.

Working in the Schrodinger picture, we will calculate the amplitude
\beq
\mathcal{A}(k,l)=\langle H,k;L,l|e^{-iH t}|0\rangle \label{akl}
\eeq
where $|0\rangle$ is our initial state
\beq
|0\rangle=\int dp_1 e^{-\frac{p_1^2}{2\sigma_s^2}}\int dp_2 e^{-\frac{p_2^2}{2\sigma_d^2}}e^{-ixp_2}|L,p_2;H,p_1\rangle.
\eeq
The state $|I,p_2;J,p_1\rangle$ consists of a $\phi_{DI}$ with momentum $p_2$ and a $\phi_{SJ}$ with momentum $p_1$, where $I$ and $J$ run over the indices $L$ and $H$.  The constants $\sigma_s$ and $\sigma_d$ are the initial wave packet momentum spreads of the source and detector, which are fixed by the experimenter.  The source is centered at the position $0$ whereas the position of the center of the detector is $x$.  Note that momentum conservation implies that, before integration over $p_1$ and $p_2$, the amplitude (\ref{akl}) is proportional to $\delta(l+q-p_1)\delta(k-q-p_2)$ where $q$ is the momentum transfer.

To second order in $H_I$, there are two possible processes with final states with no neutrinos.  First, a neutrino may travel from the source to the detector.  Second, a neutrino may travel from the detector to the source.  If we let the masses $M_{IH}$ of the $H$ particles be more massive than those $M_{IL}$ of the $L$ particles, where $I$ runs over $\{S,D\}$ then the second process will be far off shell and will have a negligible contribution for macroscopic baselines \cite{tempo}.  Therefore we will consider only the contribution from the first process.

The amplitude can be written as a sum over contributions from each neutrino mass eigenstate $\psi_1$ and $\psi_2$ with mass $m_1$ and $m_2$
\beq
\mathcal{A}(k,l)=\sum_{i=1}^2\mathcal{A}_i(k,l).
\eeq
These in turn are given by integrals over the neutrino momentum $q$ \cite{noimisura}
\bea
\mathcal{A}_i(k,l)&=&\int\frac{dq}{2\pi}\frac{F_i(q)}{c_i(q)}\expb{-\frac{(l+q)^2}{2\sigma_s^2}-\frac{(k-q)^2}{2\sigma_d^2}-ix(k-q)}\label{amp}\\
c_i(q)&=&8 e_i(q)\sqrt{E_{SL}(l)E_{SH}(l+q)E_{DL}(k-q)E_{DH}(k)}\nonumber\\
F_i&=&-\int_0^t dt_1\int_0^{t-t_1} dT \nonumber\\
&&\times \expb{-i (t_1 \mathcal{E}_0(l+q,k-q)+T\mathcal{E}_{1i}(l,k-q,q)+(t-T-t_1)\mathcal{E}_2(l,k)}.\nonumber
\eea
Here $t_1$ and $T$ are naturally interpreted as the neutrino creation time and propagation time.  We remind the reader that our experiment begins at time 0 in the state $|0\rangle$ and concludes at time $t$ when the interactions are switched off.

The on-shell energies are 
\beq
E_{\alpha I}(p)=\sqrt{M_{\alpha I}^2+p^2}\hsp
e_i(p)=\sqrt{m_i^2+p^2}
\eeq
which are summands in the eigenvales of the free Hamiltonian $H_0$ before neutrino production, during neutrino propagation and after neutrino absorption respectively
\bea
\E_0(p_1,p_2)&=&E_{SH}(p_1)+E_{DL}(p_2)\hsp
\E_{1i}(p_1,p_2,q)=E_{SL}(p_1)+E_{DL}(p_2)+e_{i}(q)\nonumber\\
\E_2(p_1,p_2)&=&E_{SL}(p_1)+E_{DH}(p_2). 
\eea
We note that, as always in a Lorentz-invariant quantum field theory, momentum and energy are exactly conserved at each vertex.  However energy is the eigenvalue of $H$.  On the other hand $\mathcal{E}_i$ is the eigenvalue of $H_0$ and so in general $\mathcal{E}_0$, $\mathcal{E}_{1i}$ and $\mathcal{E}_2$ will not be equal except when all particles are exactly on-shell.

\subsection{Saddle Point Approximation to the Amplitude}\label{saddle_point}

Only the final momentum $k$ of the detector is measured.  Once the conservation of momentum has been imposed, the only other two momenta in the problem are the neutrino momentum $q$ and the final source momentum $l$.  Once $k$ is fixed, if one demands that all particles be on-shell then $q$ and $l$ will be fixed to the on-shell values $q_i$ and $l_i$, where the $i$ index reminds the reader that these depend on the neutrino mass eigenstate $i$.  The dependence of $q_i$ and $l_i$ on $k$ will be left implicit.  On-shell the energies $\mathcal{E}$ agree
\beq
\mathcal{E}_0(l_i+q_i,k-q_i)=\mathcal{E}_{1i}(l_i,k-q_i,q_i)=\mathcal{E}_2(l_i,k)=\epsilon_i.
\eeq
 
Our particles will not be on-shell.  But they will nearly be on-shell.  Therefore we may expand the energies $\mathcal{E}$ about the on-shell value.  For example, to linear order in $q-q_i$ but zeroeth order in $l-l_i$ one finds
\beq
\mathcal{E}_0(l+q,k-q)=\epsilon_i+v_{0i}(q-q_i)\hsp
\mathcal{E}_{1i}(l,k-q,q)=\epsilon_i+v_{1i}(q-q_i)\hsp
\mathcal{E}_2(l,k)=\epsilon_i \label{qexp}
\eeq
where we have defined the on-shell velocities
\beq
v_{0i}=v_{SH,i}-v_{DL,i}=\frac{l_i+q_i}{E_{SH}(l_i+q_i)}-\frac{k-q_i}{E_{DL}(k-q_i)}\hsp
v_{1i}=v_{\psi_i}-v_{DL,i}=\frac{q_i}{e_i(q_i)}-\frac{k-q_i}{E_{DL}(k-q_i)}.
\eeq
%We have kept their dependence on the mass eigenstate $i$ implicit to avoid clutter.

This expansion of the energy allows us to expand the amplitude (\ref{amp}) and perform the integral over $q$
\bea
\mathcal{A}_i(k,l)&=&-\frac{B_i}{c_i(q_i)}\label{abc}\\
B_i&=&\int_0^t dt_1\int_0^{t-t_1} dTe^{\gamma_i}\int\frac{dq}{2\pi}\expb{-\frac{\sigma_x^2 q^2}{2}+\beta_i q}\nonumber\\
&=&\frac{1}{\sigma_x\sqrt{2\pi}}\int_0^t dt_1\int_0^{t-t_1} dT\expb{\gamma_i+\frac{\beta_i^2}{2\sigma_x^2}}\nonumber\\
\sigma_x&=&\left(\frac{1}{\sigma_s^2}+\frac{1}{\sigma_d^2}\right)^{1/2}\nonumber\\
\beta_i&=&i(\delta_i-d_i)\hsp \delta_i=x+i\left(\frac{l}{\sigma_s^2}-\frac{k}{\sigma_d^2}\right)\hsp d_i=t_1v_{0i} + T v_{1i}
\nonumber\\
\gamma_i&=&-\frac{l^2}{2\sigma_s^2}-\frac{k^2}{2\sigma_d^2}+i\left(-xk-t\epsilon_i+q_i d_i\right). 
\nonumber
\eea
To perform the $T$ integral, one need only complete the square
\bea\label{def_rho}
\int_0^{t-t_1} dT\expb{\gamma_i+\frac{\beta_i^2}{2\sigma_x^2}}&=&e^{-\rho_i}\int_0^{t-t_1} dT\expb{-\mu_i (T-T_{0i})^2}\nonumber\\
\rho_i&=&\frac{(l+q_i)^2}{2\sigma_s^2}+\frac{(k-q_i)^2}{2\sigma_d^2}+i\left(x(k-q_i)+t\epsilon_i\right)\nonumber\\
\mu_i&=&\frac{v_{1i}^2}{2\sigma_x^2}\hsp T_{0i}=\frac{\delta_i-t_1v_{0i}+i\sigma_x^2 q_i}{v_{1i}}.
\eea
Again, to avoid clutter, the dependences on $k$ are left implicit.

Now we come to the key simplification.  If $\mu$ is much larger than $1/\sqrt{t-T}$, and the imaginary part of $T_0$ is small enough, then the Gaussian on the first line is essentially a Dirac delta function
\beq
\expb{\mu_i (T-T_{0i})^2}\sim\sqrt{\frac{\pi}{\mu_i}}\delta(T-T_{0i})=\sqrt{2\pi}\frac{\sigma_x}{v_{1i}}\delta(T-T_{0i})
\eeq
and the integral gives $\sqrt{\pi/\mu}$ if Re($T_0$) is in the range of integration $[0,t-t_1]$ and otherwise gives zero
\beq
\int_0^{t-t_1} dT\expb{\mu_i (T-T_{0i})^2}=\sqrt{2\pi}\frac{\sigma_x}{v_{1i}}\theta(\Re{T_0})\theta(t-t_1-\Re{T_0})
\eeq
where $\theta$ is the Heaviside step function.  The product of step functions is nonzero whenever
\beq
0\leq x-t_1v_{0i}\leq v_{1i} (t-t_1).
\eeq
This is just the condition that the on-shell neutrino can travel as far as the detector if it is emitted at time $t_1$.   As $v_{1i}>v_{0i}$ for ultrarelativistic neutrinos, this implies
\beq
t_1\leq \frac{t v_{1i}-x}{v_{1i}-v_{0i}}.
\eeq
If the neutrino is emitted after this time, it will not arrive at the detector before it is measured at time $t$.  The $t_1$ integral is then easily computed.  So long as $x>v_{0i}t$, which means that the source and detector have not moved past one another, one finds
\bea\label{int_theta}
\int_0^t dt_1\int_0^{t-t_1} dT\expb{\mu_i (T-T_{0i})^2}&=&\sqrt{2\pi}\frac{\sigma_x}{v_{1i}}\int_0^t dt_1 \theta\left(\frac{t v_{1i}-x}{v_{1i}-v_{0i}}-t_1\right)\\
&=&\sqrt{2\pi}\frac{\sigma_x}{v_{1i}}\frac{t v_{1i}-x}{v_{1i}-v_{0i}}\theta\left(t v_{1i}-x\right).\nonumber
\eea

Putting this all together
\beq
\mathcal{A}_i(k,l)=-\frac{(t v_{1i}-x)\theta\left(t v_{1i}-x\right)}{c_i(q_i) v_{1i}(v_{1i}-v_{0i})} e^{-\rho_i}. \label{ampsad}
\eeq

\subsection{Ultrarelativistic and Small Mass Splitting Approximations}

Each $q_i$ may be obtained from the defining relation
\beq
0=\epsilon_i-\epsilon_i=\mathcal{E}_{1i}(l_i,k-q_i,q_i)-\mathcal{E}_2(l_i,k)=\sqrt{M^2_{DL}+(k-q_i)^2}+\sqrt{q_i^2+m_i^2}-\sqrt{M^2_{DH}+k^2} \label{qidef}
\eeq
which happens to be independent of $l_i$.  We will consider the leading term in two expansions.  First, the ultrarelativistic expansion is a series in $m_i^2/q_i^2$ and second we will fix an arbitrary $q_0$ and consider a power series in $q_i-q_0$.  This second expansion obviously is valid for $q_0$ close enough to $q_i$, but we will use the same expansion for every flavor $i$ and so the expansion is valid for some $q_0$ if $q_2-q_1$ is smaller than the other momenta in the problem, which we will see occurs when the mass splitting is small.  The ultrarelativistic expansion is performed first, and so $q_2-q_1$ need not be smaller than $m_i$.  

The leading term in the double expansion of Eq.~(\ref{qidef}) is
\beq
-\sqrt{M^2_{DL}+(k-q_0)^2}-q_0+\sqrt{M^2_{DH}+k^2}=\frac{m_i^2}{2q_0}+(q_i-q_0)\left[1-\frac{m_i^2}{2q_0}-\frac{k-q_0}{\sqrt{M^2_{DL}+(k-q_0)^2}}\right].
\eeq
The left hand side is independent of the flavor $i$.  Subtracting the right hand sides at two flavors $i$ and $j$ one finds
\beq
q_i-q_j=\frac{(m_j^2-m_i^2)/(2q_0)}{1-\frac{m_i^2}{2q_0}-\frac{k-q_0}{\sqrt{M^2_{DL}+(k-q_0)^2}}}\sim\frac{m_j^2-m_i^2}{2q_0}. 
\eeq
The leading term in the expansion is a good approximation if $q_0\sim q_i$ for all $i$.  In the ultrarelativistic approximation, these in turn are roughly equal to the on-shell neutrino energy, which we will call $e$.  Thus we arrive at our final expression
\beq
q_2-q_1=\frac{m_1^2-m_2^2}{2e}. \label{dq}
\eeq
%where $e$ is the on-shell neutrino energy.  As we are in the ultrarelativistic approximation, $e_1$ and $e_2$ are close, and so at leading order we may write them both as the same $e$  in Eq.~(\ref{dq}).  
This does not imply that we are approximating the neutrinos to actually have the same energy, the neutrino momenta for each mass eigenstate have been integrated over all values of $q$ and no cross-terms have been dropped.  Recall that $q_1$ and $q_2$ are just the on-shell values of the momenta.  

Similarly we will approximate
\beq
c=c_i(q_i)\hsp v_0=v_{0i}\hsp v_1=v_{1i}\hsp \epsilon=\epsilon_i. \label{apps}
\eeq
These approximations may be justified via double expansions such as that above.  In addition, given the kinematics of our process, the on-shell neutrino energies for the two eigenstates are nearly equal
\beq
e_2(q_2)-e_1(q_1)=\sqrt{q_2^2+m_2^2}-\sqrt{q_1^2+m_1^2}\sim q_2+\frac{m_2^2}{2q_2}-q_1-\frac{m_1^2}{2q_1}\sim q_2-q_1-\frac{m_1^2-m_2^2}{2e}
\eeq
which vanishes in our approximation by Eq.~(\ref{dq}).  %Now the flavor dependence of the total energy is
%\beq
%\epsilon_2-\epsilon_1=\mathcal{E}_{12}(l_2,k-q_2,q_2)-\mathcal{E}_{11}(l_1,k-q_1,q_i)
%\eeq

\subsection{The Probability Density}

The unnormalized probability density is
\beq
P(k,l)=|\mathcal{A}_1(k,l)+\mathcal{A}_2(k,l)|^2. \label{pkl}
\eeq
It may be normalized as in Ref.~\cite{noi1}, but we will not normalize it here.  It is now easily computed
\bea \label{P_kl}
P(k,l)&=&\left(\frac{(t v_{1}-x)}{c v_{1}(v_{1}-v_{0})} \right)^2\theta(tv_1-x)\label{pkl2}\\
&&\times
\left[2\expb{-\frac{(l+q_1)^2+(l+q_2)^2}{2\sigma_s^2}-\frac{(k-q_1)^2+(k-q_2)^2}{2\sigma_d^2}}\cos\left(\frac{x(m_1^2-m_2^2)}{2e}\right)\right.\nonumber\\
&&\left.
+\sum_{i=1}^2\expb{-\frac{(l+q_i)^2}{\sigma_s^2}-\frac{(k-q_i)^2}{\sigma_d^2}}\right].\nonumber
\eea
We are interested in the case in which $l$ is not observed, and so we must integrate over $l$, yielding
\bea
P(k)&=&\int dlP(k,l)=\left(\frac{(t v_1-x)}{c v_{1}(v_{1}-v_{0})} \right)^2\theta(tv_1-x)\nonumber\\
&&\times
\left[2\expb{-\frac{(k-q_1)^2+(k-q_2)^2}{2\sigma_d^2}-\left(\frac{(m_1^2-m_2^2)}{4e\sigma_s}\right)^2}\cos\left(\frac{x(m_1^2-m_2^2)}{2e}\right)\right.\nonumber\\
&&\left.+\sum_{i=1}^2\expb{-\frac{(k-q_i)^2}{\sigma_d^2}}
\right]. \label{pk}
\eea
This is the function that would be determined by an experiment which perfectly measures $k$.  

It is clearly wrong.  At large $t$ it is proportional to $t^2$.  However the source strength is constant at this leading order in perturbation theory, and so the probability $P(k)$ of having absorbed a neutrino by time $t$ should be proportional to $t$ \cite{as10}. We will see in the Sec. \ref{source_momentum} that this is an artifact of the approximations used in this section: indeed, if the contribution from the off-shell momenta of the source is taken into account, $P(k)$ grows linearly in $t$; a more rigorous proof can be found in \ref{app}.  Let us ignore this problem for the moment, as it does not affect the exponent, which is the part of interest to us.

The total, unnormalized oscillation probability can be found by integrating (\ref{pk}) over $k$, yielding
\bea
P&=&\int dk P(k)\label{p}\\
&=& 2\left(\frac{(t v_1-x)}{c v_{1}(v_{1}-v_{0})} \right)^2\theta(tv_1-x)\left[1+\expb{-\left(\frac{(m_1^2-m_2^2)}{4e}\right)^2\sigma_x^2}\cos\left(\frac{x(m_1^2-m_2^2)}{2e}\right).\right] \nonumber
\eea
Here we have assumed that $\sigma_d$ is very narrow, and so have ignored the dependence of $q_i$ on $k$.  Had we not done this, we would have found an additional source of decoherence due to the uncertain momentum $k$.  However this would not be an intrinsic source of decoherence, as an ideal detector can measure $k$ as precisely as desired given enough time.  For example, one can wait until the detector smears as closely as desired to a plane wave.  

Note that when the arguments of the exponentials are small, there is no decoherence and we recover the standard oscillation formula
\beq
P\sim 2\left(\frac{(t v_1-x)}{c v_{1}(v_{1}-v_{0})} \right)^2\theta(tv_1-x)\cos^2\left(\frac{x}{ L_{osc}}\right)\hsp
L_{osc}=\frac{4e}{m_1^2-m_2^2}.
\eeq
However, we see in Eq.~(\ref{p}) that more generally the oscillations decohere due to the factor exp$(-\sigma_x^2/L_{osc}^2)$.  This is just the usual decoherence due to an uncertain neutrino travel distance reported, for example, in Refs.~\cite{giunti93,giunti97}.

On the other hand, at the order considered in our various expansions, there is no sign of intrinsic decoherence due to the uncertainty in the neutrino energy or momentum, despite the fact that the oscillation pattern depends on the neutrino momentum and energy.  This is also despite the fact that the detector has a large momentum spread $\sigma_d$, which one might expect to wash out oscillations with a momentum difference beneath this threshold.  The reason that such oscillations are not washed out is that the on-shell condition fixes the momenta $q$ which contribute to the amplitudes to a very narrow range, much narrower than $\sigma_d$.  This counterintuitive fact is a result of the very long time integration in the definition of our amplitude, reflecting the fact that the neutrino may be produced at any time.  This long time integration forces the neutrinos to be very close to on-shell.  When we approximated the $T$ integral by that of a delta function, we effectively imposed an infinite time integration and so fixed the neutrino momenta for each value of $k$ and mass eigenstate.  That is not to say that the detector can measure the neutrino momentum $q$ more precisely than its intrinsic scatter $\sigma_d$, on the contrary it cannot tell which mass eigenstate arrived and the two mass eigenstates have very different on-shell momenta.  

\subsection{Numerical Results}

In this subsection we will test the above results numerically.  Following Ref.~\cite{noimisura} we tune the masses to optimize the sensitivity of the detector
\beq
M_{DH}=M_{SH}\hsp M_{SL}=M_{DH}(1-\epsilon)\hsp
M_{DL}=M_{SH}\left(1-\epsilon+\epsilon^2\right)
\eeq
and we then fix the parameters
\beq
M_{DH}=10\hsp \epsilon=0.1.
\eeq
In addition we fix the neutrino masses
\beq
m_1=0\hsp m_2=0.1.
\eeq

With these choices we obtain the on-shell conditions
\beq
q_1=0.9498\hsp q_2=0.9444\hsp l_1=-0.9525\hsp l_2=-0.9528 \label{guscio}
\eeq
which lead to the on-shell velocities
\beq
v_{00}=-0.0058\hsp v_{01}=-0.0069\hsp v_{10}=0.9945\hsp v_{11}=0.9883.
\eeq

\begin{figure} %[!tph]
\begin{center}
\includegraphics[width=2.5in,height=1.7in]{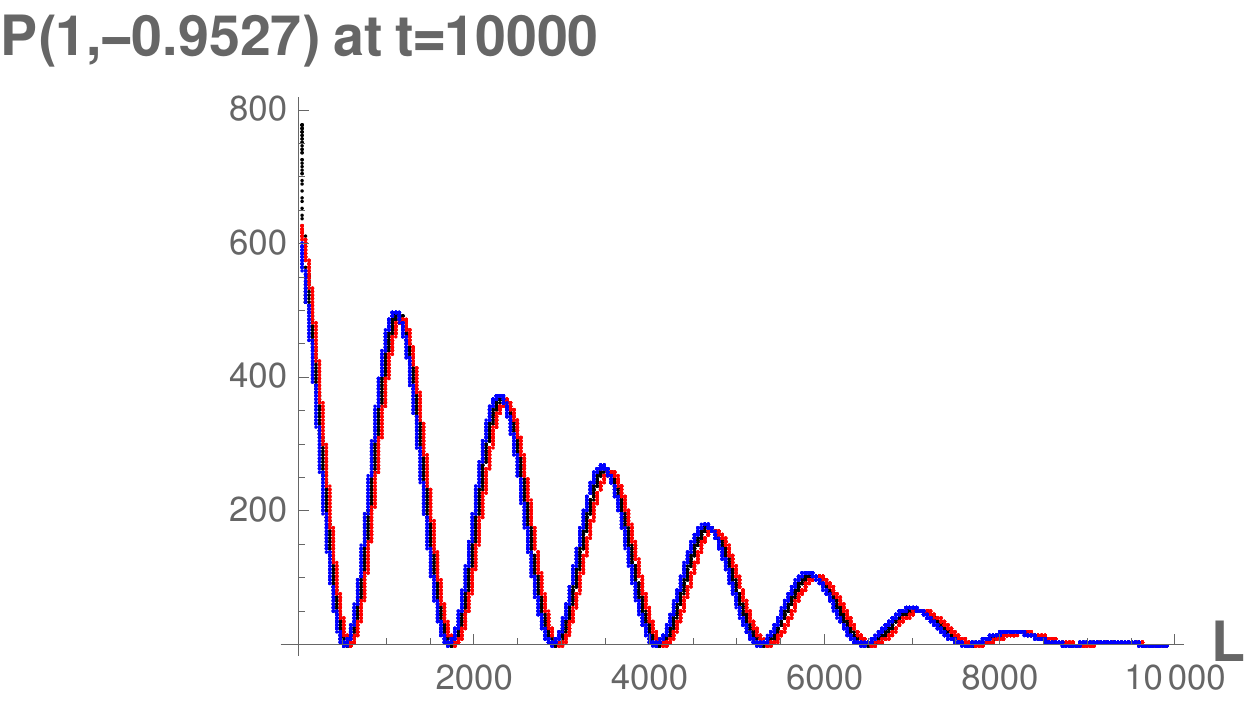}
\includegraphics[width=2.5in,height=1.7in]{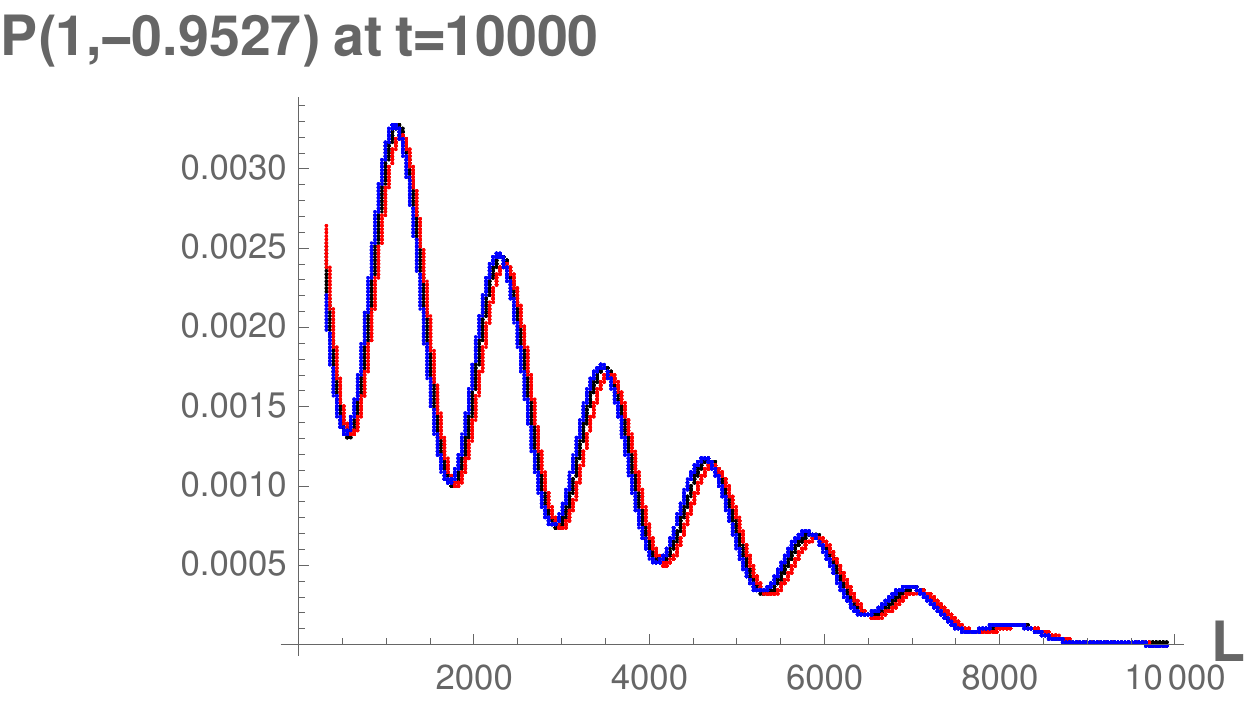}
\includegraphics[width=2.5in,height=1.7in]{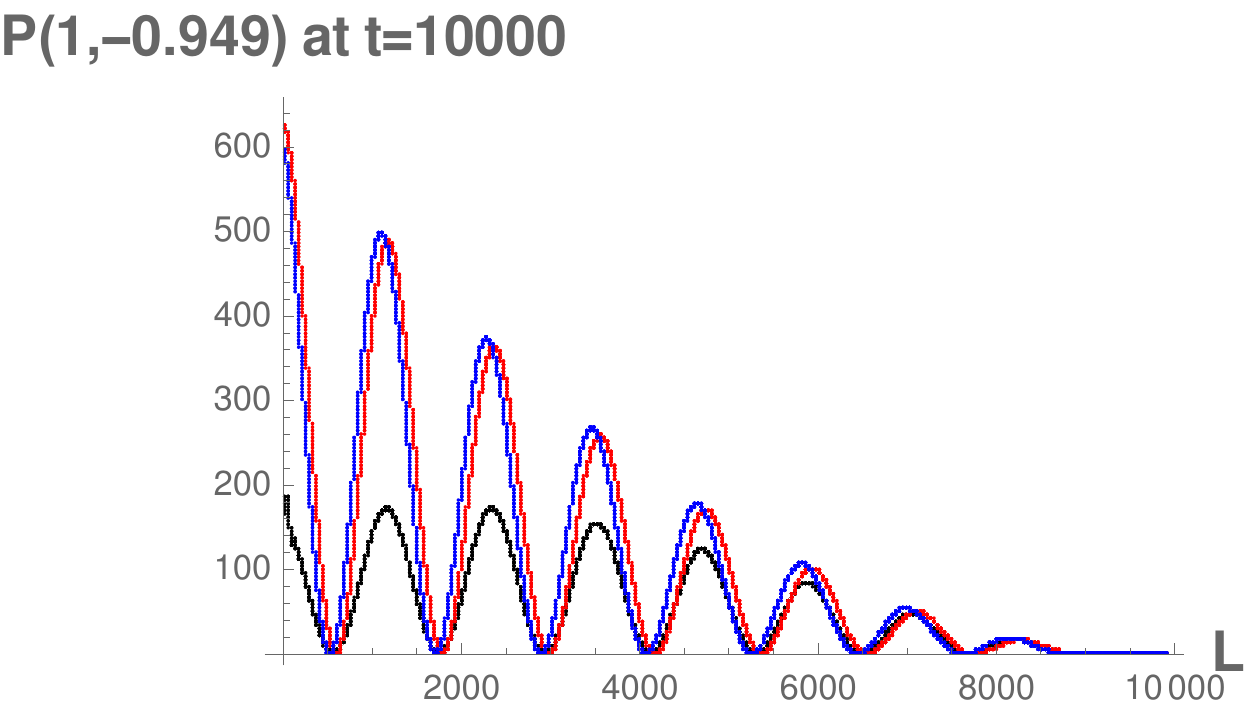}
\includegraphics[width=2.5in,height=1.7in]{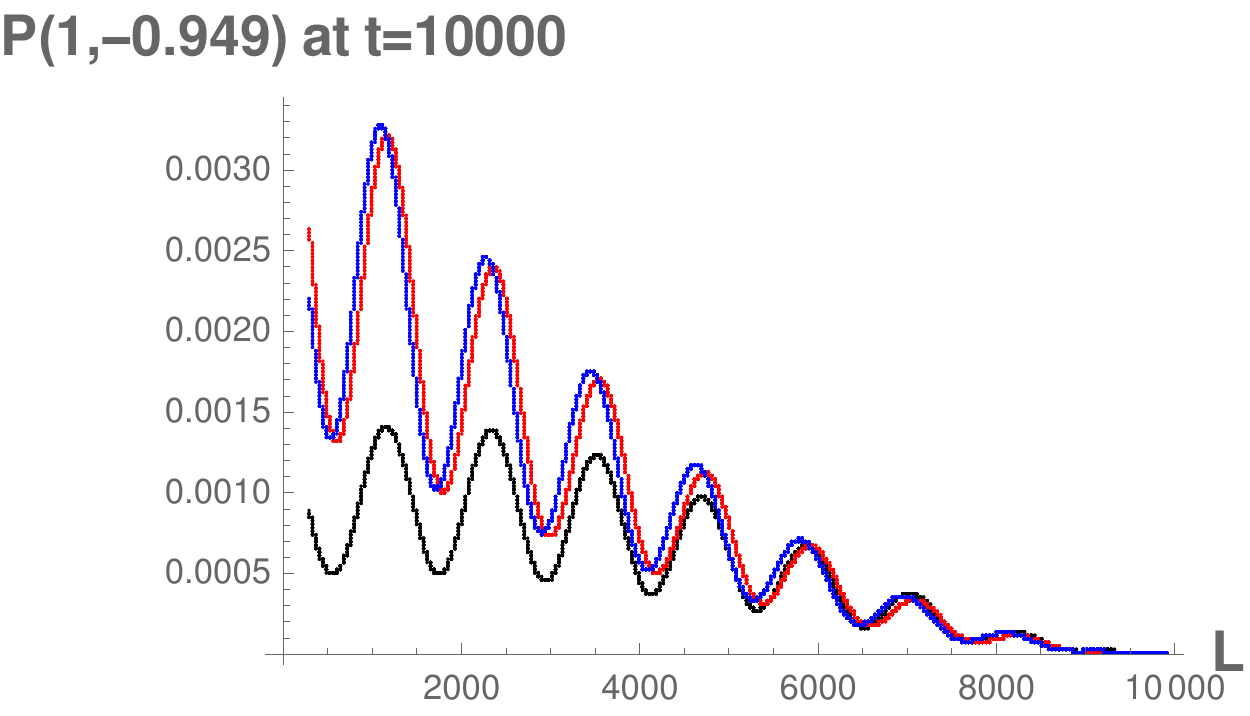}
\includegraphics[width=2.5in,height=1.7in]{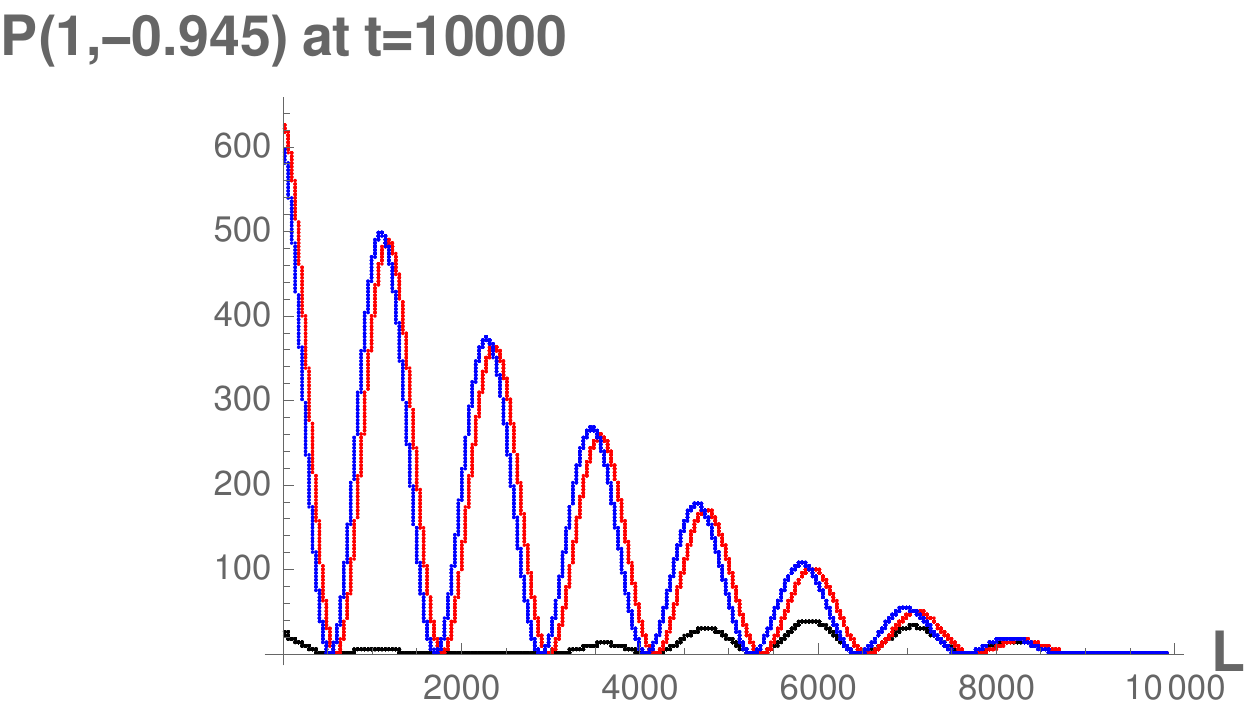}
\includegraphics[width=2.5in,height=1.7in]{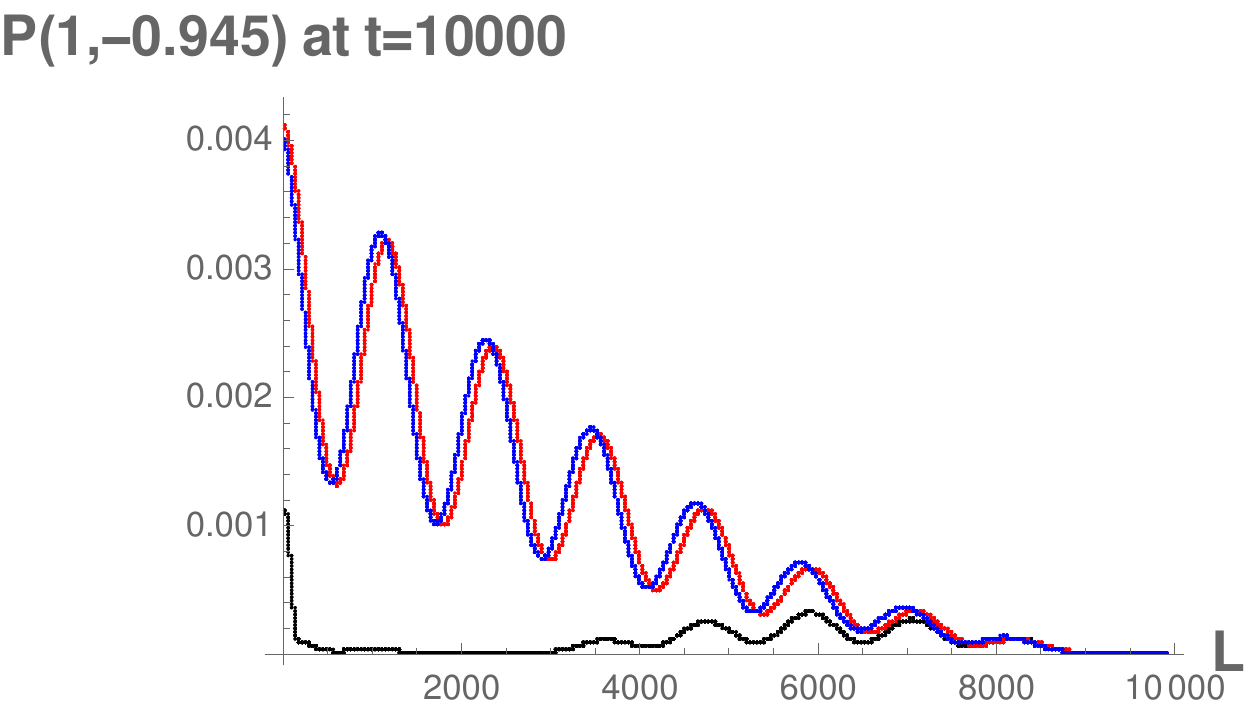}
\caption{The unnormalized probability density $P(k,l)$ at $k=1$ at various values of $l$, computed using the exact expression (\ref{amp}) and also the approximations (\ref{ampsad}) and (\ref{pkl2}).  One can see that in the top panels, where $l$ is near the on-shell value given in (\ref{guscio}), the approximations are quite accurate, but they drop too slowly when $l$ differs from the on-shell value in the lower panels.  In the left we chose $\sigma_s=\sigma_d=0.1$, on the right $\sigma_s=\sigma_d=0.015$.  Thus only on the right the source and detector sizes are of order the oscillation wavelength, and so the oscillation amplitude is reduced.}
\label{pklfig}
\end{center}
\end{figure}

At time $t=10^4$ we plot the unnormalized oscillation probability density $P(k,l)$ at $k=1$ as a function of baseline $x$ with two different sizes $1/\sigma_s$ and $1/\sigma_d$ for the source and detector, one of order the oscillation length and one much smaller.  These are plotted in three approximations.  First we use the exact second-order formula for the amplitude (\ref{amp}).  Next we use our saddle point approximation amplitude (\ref{ampsad}) and finally we use directly use our formula for the probability (\ref{pkl2}) which used the approximations (\ref{apps}).  Our results are shown in Fig.~\ref{pklfig}.    As expected, one can see that the oscillation amplitude is about 100\% when the source and detector sizes are much smaller than the oscillation wavelength, but is reduced when the sizes are comparable as a result of decoherence.  The total probability also decreases in the case of a larger source and detector as a result of our normalization of the source and detector wave functions in Eq.~(\ref{amp}).

One can see that the approximations are quite reliable when $l$ is close to the on-shell values $l_i$ given in Eq.~(\ref{guscio}), but do not capture the correct fall-off when $l-l_i$ is increases.  This is caused by our very crude expansion of the energy in Eq.~(\ref{qexp}), which was to zeroeth order in $l-l_i$.  We will now expand it to first order in $l-l_i$.

\subsection{The Source Momentum}\label{source_momentum}

To first order in $l-l_i$ the energy may be expanded
%The problem is our very crude expansion of the energy in Eq.~(\ref{qexp}), which was to zeroeth order in $l-l_i$.  We will now expand to first to order in $l-l_i$
\bea
\mathcal{E}_0(l+q,k-q)&=&\epsilon_i+u_{0i}(l-l_i)+v_{0i}(q-q_i)\label{lexp}\\
\mathcal{E}_{1i}(l,k-q,q)&=&\epsilon_i+u_{1i}(l-l_i)+v_{1i}(q-q_i)\hsp
\mathcal{E}_2(l,k)=\epsilon_i+u_{2i}(l-l_i) \nonumber
\eea
where the new on-shell velocities are
\beq \label{def_u}
u_{0i}=v_{SH,i}=\frac{l_i+q_i}{E_{SH}(l_i+q_i)}\hsp
u_{1i}=u_{2i}=v_{SL,i}=\frac{l_i}{E_{SL}(l_i)}.
\eeq
Note that at $l=l_i$ the two expansions (\ref{lexp}) and (\ref{qexp}) agree, and so one expects the probability density $P(k,l_i)$ in (\ref{pkl}) to be correct, as can be seen in the top panels of Fig.~\ref{pklfig}.

The calculation above proceeds similarly to the zeroeth order case studied above.  As $u_{1i}=u_{2i}$, the on-shell neutrino propagation time $\Re{T_0}$ is unaffected and Eq.~(\ref{abc}) is also the same except for a shift in $\gamma_i$
\beq
\gamma_i\longrightarrow \gamma_i\p =\gamma_i-i(l-l_i)(t_1 u_{0i}+(t-t_1)u_{2i}).  \label{lpeso}
\eeq
As the $T$-dependent terms are unaffected, the $T$ integral leads to the same Heaviside step function as before, but now with a $t_1$-dependent phase $e^{\gamma_i\p}$ from (\ref{lpeso}).  As a result, the $t_1$ integral is no longer trivial.  

This is to be expected on physical grounds.  Holding $t_1$ fixed and varying $T$, one multiplies the amplitude by a phase $e^{iT_0(\mathcal{E}_2-\mathcal{E}_1)}$ which is independent of $l$.  The $T$ integral therefore fixes $q$ to be near its on-shell value but does not constrain $l$.  On the other hand, varying $t_1$ one multiples by phase $e^{i t_1(\mathcal{E}_1-\mathcal{E}_0)}$ which depends on $l$.  Therefore it is the $t_1$ integral which forces $l$ to be on-shell.

Incorporating the new phase from (\ref{lpeso}), together with the old step function, the $t_1$ integral is
\beq\label{amp_l}
\int_0^t dt_1 \theta(t-t_1-\Re{T_0}) e^{-it_1(l-l_i)(u_{0i}-u_{2i})}=i\frac{ e^{i\alpha(l-l_i)}-1}{(l-l_i)(u_{0i}-u_{2i})}
\eeq
where
\beq\label{def_alpha}
\alpha=\frac{(u_{0i}-u_{2i})(tv_{1i}-x)}{v_{1i}-v_{0i}}.
\eeq
This new factor multiplies $P(k,l)$.  It tends to unity if
\beq
\left|l-l_i\right|<<\left|\frac{v_{1i}-v_{0i}}{(tv_{1i}-x)(u_{0i}-u_{2i})}\right| \label{nufact}
\eeq
but approaches zero for higher values.  One integrates $P(k,l)$ over l to obtain $P(k)$.  The restricted range (\ref{nufact}) causes $P(k)$ to lose one power of $\alpha$, or equivalently one power of $tv_1-x$.  As a result, now $P(k)$ grows only linearly with respect to $tv_1-x$, not quadratically as in Eq.~(\ref{pk}). In \ref{app}, where we solve exactly the integral and calculate the expression for $P(k)$, we prove this claim. This linear dependence was shown, in Ref.~\cite{as10}, to be a general consequence of energy conservation. 

Eq.~(\ref{nufact}) has a simple, physical interpretation.  $u_0-u_2$ is the velocity recoil of the source when it admits the neutrino.  $t-x/v_1$ is the time difference between the first allowed emission time $t_1=0$ and the last time at which the neutrino may be emitted and arrive at the detector by time $t$.  Therefore the product $(u_0-u_2)(t-x/v_1)$ is the size of range of possible centroids of the source when the neutrino is emitted.  Eq.~(\ref{nufact}) is then just the uncertainty principle, the source momentum $l$ cannot be constrained more tightly than this range in source positions.  Interestingly the intrinsic source size, $1/\sigma_s$, does not play any role in this manifestation of the uncertainty principle.  Of course were $\sigma_s$ too small, then $\sigma_x$ would be large and so the step function approximation of the $T$ integral would be invalid.  However this corresponds to the case in which the run time of the experiment is comparable to the uncertainty in the neutrino travel time, which is never realized in practice.    The fact that the momentum uncertainties depend on the velocity recoil and not the momentum smearing of the source and detector is in contradiction with the usual intuition \cite{mcdonald}, but it is a robust implication of our dynamical treatment of the source and detector fields.

%Note that if the uncertainty in $l$ resulting from the recoil is larger than $\sigma_x$, then neutrinos emitted at different times will yield source particles at locations so different that the final states no longer overlap.  As a result, neutrino emissions at sufficiently different times would no longer be summed coherently.  This effect is incorporated in our formalism via the Gaussian damping terms in (\ref{amp}).  It implies that $\sigma_x$ provides a kind of {\it{upper}} bound on the momentum uncertainty.  

\section{Maximal Coherence Length?} \label{maxsez}

In Ref.~\cite{giuntimax}, as in our Sec.~\ref{dynsez}, the authors consider a 1+1 dimensional model of scalar neutrinos in the Schrodinger picture of quantum field theory.  The neutrinos are described by Gaussian wave packets, characterized by a single spatial width $\sigma_x$, and the source and detector particles are not explicitly considered.  The authors find the same two contributions to decoherence as in Ref.~\cite{giunti97}.   Decoherence due to the baseline uncertainty provides an upper bound on $\sigma_x$ beyond which oscillations cannot be observed, while that due to the momentum uncertainty of the wave packet provides a lower bound.  Oscillations cannot be measured for any $\sigma_x$ if the upper bound is less than the lower bound, which the authors find always occurs beyond some distance $L_{\rm{max}}$ which depends only on the neutrino mass splitting and energy
\beq
L_{\rm{max}}=\frac{16\pi^2 e^3}{(m_2^2-m_1^2)^2}.
\eeq

Needless to say, our results are in stark contradiction with those of Ref.~\cite{giuntimax}, even those which are found in Sec.~\ref{extsez} using an external wave packet model\footnote{Ref.~\cite{beuthe} also finds that there is no maximum coherence length in an external wave packet model.} introduced by the same authors a few months earlier in Ref.~\cite{giunti97}.  This contradiction stems from the fact that we, like Ref.~\cite{grimus}, do not find any intrinsic source of decoherence to the momentum uncertainty, and so we have no lower bound on $\sigma_x$.  This, in turn, is due to the entanglement of the detector with the neutrino, which is not considered in Ref.~\cite{giuntimax} as the detector wave function is not considered explicitly in that work.  In our case, neutrinos with different emission times $t_1$ are summed coherently, while $t_1$ varies over a macroscopic interval whose size is of order $t$.  In the language of wave packets, this would naively imply that our wave packet width $\sigma_w$ is essentially infinite (of order $ct$, which is literally astronomical) as one cannot say where the neutrino is located.  Wave packet intuition would then suggest that neutrino oscillations should be washed out, as they are damped by $\expb{(\sigma_x/L_{osc})^2}$.  However this is not the case, as the $\sigma_w$ mentioned here is the uncertainty in the neutrino's position at a fixed time, it is not the uncertainty in the baseline $\sigma_x$.  The uncertainty in the baseline $\sigma_x$ is bounded by the size of the source and detector, which is certainly finite and much smaller than $ct$.  Considering the neutrino in isolation, as in Ref.~\cite{giuntimax}, one apparently cannot distinguish $\sigma_x$ from $\sigma_w$.  It is their identification which leads to an apparent maximum coherence length. 

\begin{figure} %[!tph]
\begin{center}
\includegraphics[width=3.75in,height=2.55in]{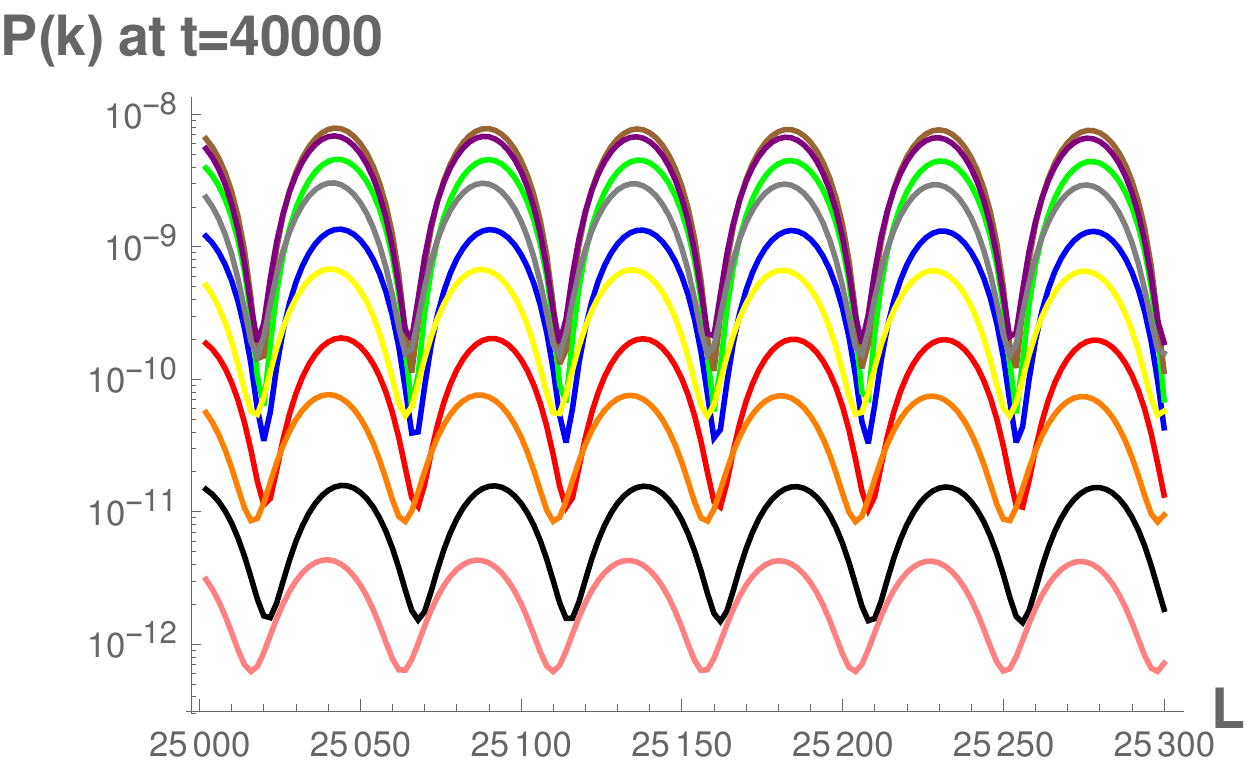}
\caption{The unnormalized probability density $P(k)$ at $k=0$ (black), 0.2 (red), 0.4 (blue), 0.6 (green), 0.8 (brown), 1.0 (purple), 1.2 (grey), 1.4 (yellow), 1.6 (orange) and 1.8 (pink)
 is plotted as a function of the baseline.  It is obtained from Eq.~(\ref{amp}) via numerical integration, and so does not use the approximations introduced in this note.  One can see oscillations well beyond the putative maximum $L_{\rm{max}}\sim 3\times 10^3$.}
\label{lfig}
\end{center}
\end{figure}

The absence of a maximum coherence length does not appear to be an artifact of the approximations in this note.  We have checked numerically that the full probability density obtained from (\ref{amp}) manifests coherent oscillations for an arbitrary time.  For example, consider
\bea
&\sigma_d=\sigma_s=0.5\hsp
\epsilon=10^{-4}\hsp
m_{SH}=m_{DH}=10^4\hsp m_{SL}=m_{DH}(1-\epsilon)\nonumber\\
&m_{DL}=m_{SH}\left(1-\epsilon+\epsilon^2\right)\hsp
m_1=0.5\hsp m_2=0.
\eea
The kinematics dictates that the neutrino energy will be about $1$ when $k\sim1$ and so
\beq
L_{\rm{max}}\sim 256\pi^2\sim 3\times 10^3.
\eeq
In Fig.~\ref{lfig} we plot the unnormalized oscillation probability density $P(k)$ at time $t=4\times 10^4$.   One can observe oscillations with amplitude of order unity at $L>>L_{\rm{max}}$.

%Note that we have considered more massive source and detector particles in this section then in the previous section.  Critically, the recoil is much smaller.  This is essential for several reasons.  First of all, it prevents the source and detector from spreading by more than an oscillation length.  This is irrelevant in actual experiments, where the source and detector are generally fixed.  However the small recoil also implies that the final states of the source resulting from distinct neutrino emission times will have significant overlap, and so these distinct emission times are summed coherently.  Only after summing coherently over emission times are the neutrinos nearly on-shell, and so this is essential for the narrow range of contributing momenta $q$, and so for the neutrino oscillations.  Had values of $q$ differing from $q_0$ by of order $\sigma_s$ or $\sigma_d$ contributed significantly to the amplitude, they would have washed out the oscillations as predicted in Ref.~\cite{giuntimax}.

\section{Remarks}

We have found that in an idealized setting, with no environmental interactions, there is no intrinsic decoherence due to the uncertain neutrino momentum.  Of course in a real detector there will an experimental imprecision that will lead to decoherence, which can easily by found by folding the energy resolution into Eq.~(\ref{pk}).  

A key step in our calculation was the replacement of our Gaussian integration over $T$ with the integral of a delta function.  This approximation seems to be easily justified in experimental setups that can be realized with present technology.  It would require an incredible time resolution to actually probe the shape of this Gaussian.  However, could it be done, the oscillations measured would not obey the usual formula.  Instead, one would observe that in a very short time window after the neutrinos arrive, they have not yet oscillated.  The neutrino detection probability actually then decreases with time at the oscillation minima, as the oscillations turn on.  This is due to destructive interference in the time integrals in our amplitude.  This interesting, but probably unobservable phenomenon, will be the subject of our next project. 

Our result is a consequence of the following argument.  The initial momentum spread of the source and detector are not in fact lower bounds on the momentum uncertainty of the neutrino.  On the contrary, in the 2-body interactions considered here, a precise determination of the final momentum of the detector is sufficient to determine the momenta of all other particles in the problem if they are on-shell.  They will of course not be exactly on-shell, but the deviation from the on-shell momentum can be much smaller than the initial momentum smearing.  %of order the inverse run time of the experiment.  This scale is much smaller than the intrinsic momentum smearing of the initial state, and so the momenta are determined much more accurately than one would expect based on this initial smearing. 

The precise enforcement of the on-shell condition is a result of the fact that the amplitude is a coherent integral over the entire runtime of the experiment, as we have not considered environmental interactions which ruin this coherence.  The final state is independent of the neutrino emission time, as is obvious when working in the basis $|H,k;L,l\rangle$ of definite momenta $k$ and $l$ as the emission time affects neither $k$ nor $l$.  As a result, processes in which neutrinos are emitted at very different times lead to the same final state, and so are added coherently.  Therefore neutrino mass eigenstates which arrive at the same time but are nonetheless coherent were emitted at different times and then coalesced following the paradigm of Ref.~\cite{mcgreevy}.

\appendix
\section{Contribution from source momentum}
\label{app}
Here we will explicitly calculate the amplitude taking into account also the off-shell momentum of the source, as discussed generally in Sec. \ref{source_momentum}. We will use the same approximations employed in the previous sections, namely we will neglect the dependence of the velocities, the neutrino energy and the numerical factor $c(q)$ on the neutrino mass eigenstate following Eq. (\ref{apps}); this will be true also for the velocities $u_0$, $u_1$ and $u_2$ defined in Eq. (\ref{def_u}).  Moreover we will also consider $l_i=l_0 \quad \forall i$: this can be justified noticing that, while $q_i-q_j\simeq(m_i^2-m_j^2)/2e$, $l_i-l_j\simeq(m_i^2-m_j^2)/2m_{SH}$, {\it i.e.} the difference between the on-shell momenta of the source considering two different neutrino mass eigenstates is suppressed by an additional factor of $m_{SH}$. From Eq. (\ref{amp_l}) we have
\beq
\mathcal{A}_i(k,l)=\frac{ie^{-\rho_i}}{c(q)v_1(u_0-u_2)}\frac{e^{-itu_2\Delta l}(1-e^{i\alpha \Delta l})}{\Delta l}\theta(tv_1-x)
\eeq
where $\rho_i$ and $\alpha$ are defined as in Eq. (\ref{def_rho}) and Eq. (\ref{def_alpha}), respectively, and $\Delta l=l-l_0$. Following the same procedure that lead to Eq. (\ref{P_kl}) we can write
\beq
P(k,l)=\frac{2}{(c(q)v_1(u_0-u_2))^2}\frac{1-\textrm{cos}(\alpha \Delta l)}{\Delta l^2}P_{nl}(k,l)
\eeq
where $P_{nl}(k,l)$is proportional to Eq. (\ref{P_kl}), up to some multiplicative factors that do not depend on l. We can write
\beq
P_{nl}(k,l)=\sum_{j=1}^3\xi_je^{-(\Delta l+\delta q_j)^2/2\sigma_j^2}
\eeq
where $\delta q_j=l_0+\tilde{q}_j$ and
\bea
\tilde{q}_j=&&\left(q_1,q_2,\frac{q_1+q_2}{2}\right)\nonumber\\
\sigma_j^2=&&\left(\sigma_s^2,\sigma_s^2,\sigma_s^2/2\right)\nonumber\\
\xi_j=&&\left(\textrm{Exp}\left[-\frac{(k-\tilde{q}_1)^2}{2\sigma_d^2}\right],\textrm{Exp}\left[-\frac{(k-\tilde{q}_2)^2}{2\sigma_d^2}\right],\textrm{Exp}\left[-\frac{(k-\tilde{q}_3)^2}{\sigma_d^2}-\frac{(q_1-q_2)^2}{\sigma_x^2}\right]\textrm{cos}\left(\frac{x(m_1^2-m_2^2)}{2e}\right)\right).\nonumber
\eea

In order to compute $P(k)$ we need to integrate over $\Delta l$ (technically speaking the integral is over $l$, however it differs from $\Delta l$ simply by a shift, so the two integrals are equivalent). We define
\beq\label{intI}
I_j(\alpha)=\int \textrm{d}(\Delta l) \frac{e^{-(\Delta l+\delta q_j)^2/2\sigma_j^2}(1-\textrm{cos}(\alpha \Delta l))}{\Delta l^2}.
\eeq
We differentiate twice with respect to $\alpha$, obtaining
\beq
I''_j(\alpha)=-\int \textrm{d}(\Delta l)(e^{-(\Delta l+\delta q_j)^2/2\sigma_j^2}(1-\textrm{cos}(\alpha \Delta l)))=\sqrt{2\pi}\sigma_je^{-\alpha^2\sigma_j^2/2}\textrm{cos}(\alpha\delta q_j)
\eeq
Solving the differential equation with respect to $\alpha$ we have
\bea
I_j(\alpha)=&&c_0+c_1\alpha+\frac{\sqrt{2\pi}\sigma_je^{-\alpha^2\sigma_j^2/2}\textrm{cos}(\alpha\delta q_j)}{\sigma_j^2}+\nonumber\\&&
\frac{e^{-\frac{\delta q_j^2}{2 \sigma_j^2}} \left(\pi  \left(\alpha\sigma_j^2-i \delta q_j\right) \textrm{Erf}\left(\frac{\alpha\sigma_j^2-i \delta q_j}{\sqrt{2} \sigma_j}\right)-\pi  \left(\delta q_j-i \alpha\sigma_j^2\right) \textrm{Erfi}\left(\frac{\delta q_j-i \alpha\sigma_j^2}{\sqrt{2} \sigma_j}\right)\right)}{2 \sigma_j^3}.
\eea
The coefficients $c_0$ and $c_1$ can be easily determined using the following argument
\begin{itemize}
\item Since the integrand in Eq. (\ref{intI}) is invariant under $\alpha\rightarrow -\alpha$, only even powers of $\alpha$ can appear, hence $c_1=0$
\item $c_0$ can be determined imposing $I_j(0)=0$, since the integrand in (\ref{intI}) identically vanishes if $\alpha=0$.
\end{itemize}
We then have
\bea
I_j(\alpha)=&&\frac{\sqrt{\pi } \left(2 \delta q_j F\left(\frac{\delta q_j}{\sqrt{2} \sigma_j}\right)-\sqrt{2} \sigma_j\right)}{\sigma_j^3}+\nonumber\\&&
\frac{e^{-\frac{\delta q_j^2}{2 \sigma_j^2}} \left(\pi  \left(\alpha\sigma_j^2-i \delta q_j\right) \textrm{Erf}\left(\frac{\alpha\sigma_j^2-i \delta q_j}{\sqrt{2} \sigma_j}\right)-\pi  \left(\delta q_j-i \alpha\sigma_j^2\right) \textrm{Erfi}\left(\frac{\delta q_j-i \alpha\sigma_j^2}{\sqrt{2} \sigma_j}\right)\right)}{2 \sigma_j^3}
\eea
where $F(x)$ is the Dawson function. In the limit $\alpha\rightarrow\infty$ we obtain
\beq
I_j(\alpha)=\frac{  \pi  \alpha  \sigma^2  e^{-\frac{\delta q_j^2}{2 \sigma ^2}}-\sqrt{2 \pi }\sigma+2 \sqrt{\pi } \delta q_j F\left(\frac{\delta q_j}{\sqrt{2} \sigma }\right)}{\sigma ^3}+\mathcal{O}(\frac{1}{\alpha}).
\eeq
Notice that the leading term scales like $\alpha$, which is proportional to $tv_1-x$, as was claimed in Sec. \ref{source_momentum}.

\section* {Acknowledgement}

\noindent
EC thanks Yoshio Kitadono for the useful discussions and suggestions. 
JE is supported by the CAS Key Research Program of Frontier Sciences grant QYZDY-SSW-SLH006 and the NSFC MianShang grants 11875296 and 11675223.  EC is supported by NSFC Grant No. 11605247, and by the Chinese Academy of Sciences Presidents International Fellowship Initiative Grant No. 2015PM063.  JE and EC also thank the Recruitment Program of High-end Foreign Experts for support.

\end{document}